\documentclass[aps,prl, showpacs, twocolumn, superscriptaddress,reprint]{revtex4-1}
\usepackage{graphicx}
\usepackage{bm}
\usepackage{gensymb}

\begin{document}

\title{Optical excitation of single- and multi-mode magnetization precession in Galfenol nanolayers}

\author{A. V. Scherbakov}
\affiliation{Experimentelle Physik 2, Technische Universit\"at Dortmund, D-44227 Dortmund, Germany}
\affiliation{Ioffe Institute, Russian Academy of Science, 194021 St.Petersburg, Russia}

\author{A. P. Danilov}
\affiliation{Experimentelle Physik 2, Technische Universit\"at Dortmund, D-44227 Dortmund, Germany}

\author{F. Godejohann}
\affiliation{Experimentelle Physik 2, Technische Universit\"at Dortmund, D-44227 Dortmund, Germany}

\author{T. L. Linnik}
\affiliation{Department of Theoretical Physics, V.E. Lashkaryov Institute of Semiconductor Physics, 03028 Kyiv, Ukraine}

\author{B. A. Glavin}
\affiliation{Department of Theoretical Physics, V.E. Lashkaryov Institute of Semiconductor Physics, 03028 Kyiv, Ukraine}

\author{L.~A.~Shelukhin}
\affiliation{Ioffe Institute, Russian Academy of Science, 194021 St.Petersburg, Russia}

\author{D. P. Pattnaik}
\affiliation{School of Physics and Astronomy, University of Nottingham, Nottingham NG7 2RD, UK}

\author{M. Wang}
\affiliation{School of Physics and Astronomy, University of Nottingham, Nottingham NG7 2RD, UK}

\author{A.~W.~Rushforth}
\affiliation{School of Physics and Astronomy, University of Nottingham, Nottingham NG7 2RD, UK}

\author{D. R. Yakovlev}
\affiliation{Experimentelle Physik 2, Technische Universit\"at Dortmund, D-44227 Dortmund, Germany}
\affiliation{Ioffe Institute, Russian Academy of Science, 194021 St.Petersburg, Russia}

\author{A. V. Akimov}
\affiliation{School of Physics and Astronomy, University of Nottingham, Nottingham NG7 2RD, UK}

\author{M. Bayer}
\affiliation{Experimentelle Physik 2, Technische Universit\"at Dortmund, D-44227 Dortmund, Germany}
\affiliation{Ioffe Institute, Russian Academy of Science, 194021 St.Petersburg, Russia}

\begin{abstract}
We demonstrate a variety of precessional responses of the magnetization to ultrafast optical excitation in nanolayers of Galfenol (Fe,Ga), which is a ferromagnetic material with large saturation magnetization and enhanced magnetostriction. The particular properties of Galfenol, including cubic magnetic anisotropy and weak damping, allow us to detect up to 6 magnon modes in a 120nm layer, and a single mode with effective damping $\alpha_{eff}=0.005$ and frequency up to 100 GHz in a 4-nm layer. This is the highest frequency observed to date in time-resolved experiments with metallic ferromagnets. We predict that detection of magnetisation precession approaching THz frequencies should be possible with Galfenol nanolayers.\end{abstract}

\maketitle

Within the last decade magnetization precession has become an actively exploited tool in nanoscale magnetism. The precessing magnetization of a ferromagnet is an effective, tunable and nanoscopic source of microwave signals of various types. Generation of microwave magnetic fields by precessing magnetization is already implemented in magnetic storage technology such as microwave assisted magnetic recording (MAMR) \cite{MAMR} by means of spin-torque nano-oscillators \cite{STO}. Spin waves or magnons, i.e. the waves of precessing magnetization, are information carriers and encoders in magnon spintronics \cite{Magnonics} aimed to substitute conventional CMOS technology. The precessing magnetization is also an effective tool to generate a pure spin current in a nonmagnetic material by means of spin pumping \cite{Spumping}.

The common way to excite magnetization precession in a ferromagnet is the technique of ferromagnetic resonance (FMR). A monochromatic microwave magnetic field drives the magnetization precession, the frequency of which is tuned into resonance with the microwaves by an external magnetic field. This technique, which can provide comprehensive information about the main precession parameters, is not adaptable for practical use with nanostructures due to the need to use bulky electromagnetic resonators and waveguides. An alternative approach is broad-band excitation induced by dc-current \cite{byCurrent}, picosecond magnetic field pulses \cite{byField} and ultrashort laser \cite{byLaser} and strain \cite{byStrain} pulses. In those cases the parameters of the excited magnetization precession, i.e. the spectral content, lifetime, spatial distribution and their dependences on external magnetic field, are determined by the properties of the ferromagnetic material and the design of the nanostructure \cite{DynamicsBook}. The ability to control these dynamical parameters is of crucial importance for nanoscale magnetic applications. For practical use, an ideal combination of dynamical parameters includes a tunable and narrow spectral band in the GHz and THz frequency ranges; large saturation magnetization and high precession amplitude for high microwave power; and ultrafast triggering for high-frequency modulation. Achieving this combination has been an unmet challenge until now. High precession frequency, $f\gg10$~GHz, can be reached by using ferrimagnetic materials \cite{Ferri1,Ferri2}, but the weak net magnetization limits their functionality. In the case of metallic ferromagnets with large net magnetization, the direct way to achieve high frequency precession is to apply a strong external magnetic field, \textbf{B}, which, however, drastically decreases the precession amplitude. Earlier experiments on the excitation of magnetization precession in metallic ferromagnets by femtosecond optical pulses \cite{byLaser,Opt1,Opt2,Opt3,Opt4,Opt5,Opt6,Opt7}, i.e. the fastest method of launching precession, report also high values of the effective damping coefficient $\alpha_{eff}=(2\pi\tau f)^{-1}>0.01$ ($\tau$ is the precession decay time). Thus, the excitation and detection of sub-THz narrow band precession in metallic ferromagnets remains extremely challenging.

In the present letter, we report the results of ultrafast magneto-optical experiments with nanolayers of (Fe,Ga), i.e. Galfenol. This metallic ferromagnet with large net magnetization is considered as a prospective material for microwave spintronics due to the narrow ferromagnetic resonance \cite{FeGaAp1, FeGaAp2} and enhanced magnetostriction \cite{SmartReview}, which allows manipulation of the magnetization direction and precession frequency by applying stress, i.e. without changing the external magnetic field \cite{FeGaAp1, FeGaSwitch}. Our study extends significantly the application potential of Galfenol. We show that in a Galfenol layer with a thickness of several nanometers, the femtosecond optical excitation leads to the generation of single-mode magnetization precession with frequency $f>100$~GHz and large amplitude. Despite the strong interaction between the magnetization and the lattice, we observe a weak damping of precession with $\alpha_{eff}\approx0.005$. Thus, we demonstrate the possibility to achieve the desirable combination of sub-THz magnetization precession with large amplitude and tunable narrow spectral band. Moreover, we show that, depending on the nanolayer thickness, we can excite multi- or single-mode magnetization precession: in a thick 120-nm Galfenol layer we observe multimode precession and resolve up to 6 precessional localized magnon modes. This allows control of the precession spectral content and spatial profile by adjusting the film thickness and excitation regime.

 The samples studied are four Fe$_{0.81}$Ga$_{0.19}$ nanolayers with thicknesses $d$= 4, 5, 20 and 120 nm grown by magnetron sputtering on (001) semi-insulating GaAs substrates and covered by a 3-nm Al or Cr cap layer to prevent oxidation. A 150-nm thick SiO$_2$ cap was deposited on the Galfenol layers with a thickness $\leq$20 nm for amplification of the magnetooptical Kerr effect \cite{Si02}. Room temperature experiments were carried out with an external magnetic field \textbf{B} applied in the layer plane. The in-plane direction of \textbf{B} is defined by the azimuthal angle $\varphi_B$ [see the inset in Fig. 1(a)]. In all studied layers the easy axes of magnetization are in the layer plane and close to the [100]/[010] crystallographic directions, while the hard axes lie along the [110] and [1$\bar{1}$0] diagonals. All nanolayers possess a weak uniaxial in-plane anisotropy, which is typical for thin Galfenol films on GaAs substrates \cite{FeGaSwitch}. We have checked that the SiO${_2}$ cap does not affect the anisotropy parameters of the layers.

The magnetization precession was excited by 150-fs pump pulses from a mode-locked Erbium-doped ring fiber laser (80 MHz repetition rate, 1050 nm wavelength). The pump beam, focused to a spot of 20 $\mu$m diameter with an energy density of $\approx1$ mJ/cm$^2$, launched the magnetization precession by ultrafast changes of the magnetic anisotropy altered by the optically-induced heating \cite{Kats}. The magnetization response was monitored using 150-fs linearly polarized probe pulses of 780-nm wavelength from another ring-fiber laser oscillator focused to a 5$\mu$m spot in the center of the pump beam. For monitoring the time evolution of the magnetization precession, we utilized the transient magneto-optical Kerr effect (TMOKE) and detected the rotation of polarization of the probe beam reflected from the (Fe,Ga) layer. In this detection scheme the signal is proportional to the changes of the magnetization projection $\Delta M_z$, where $z$ is the normal to the (Fe,Ga) layer.  The temporal resolution was achieved by means of an Asynchronous Optical Sampling System (ASOPS) \cite{Dekorsy}. The pump and probe oscillators were locked with a frequency offset of 800 Hz. In combination with the 80-MHz repetition rate, it allows measurement of the time-resolved signal in a time window of 12.5 ns with time resolution limited by the probe pulse duration.

For the measurements at magnetic fields $B>1$~T, the samples were mounted in an optical cryostat with a superconducting solenoid. In this case, the temperature of the sample was 150 K. The source of the laser pulses was a regenerative amplifier RegA (wavelength 800 nm, repetition rate 100 kHz) and a standard scanning delay line was used to monitor the temporal evolution of the magnetization.

\begin{figure}
 \includegraphics[scale=0.4]{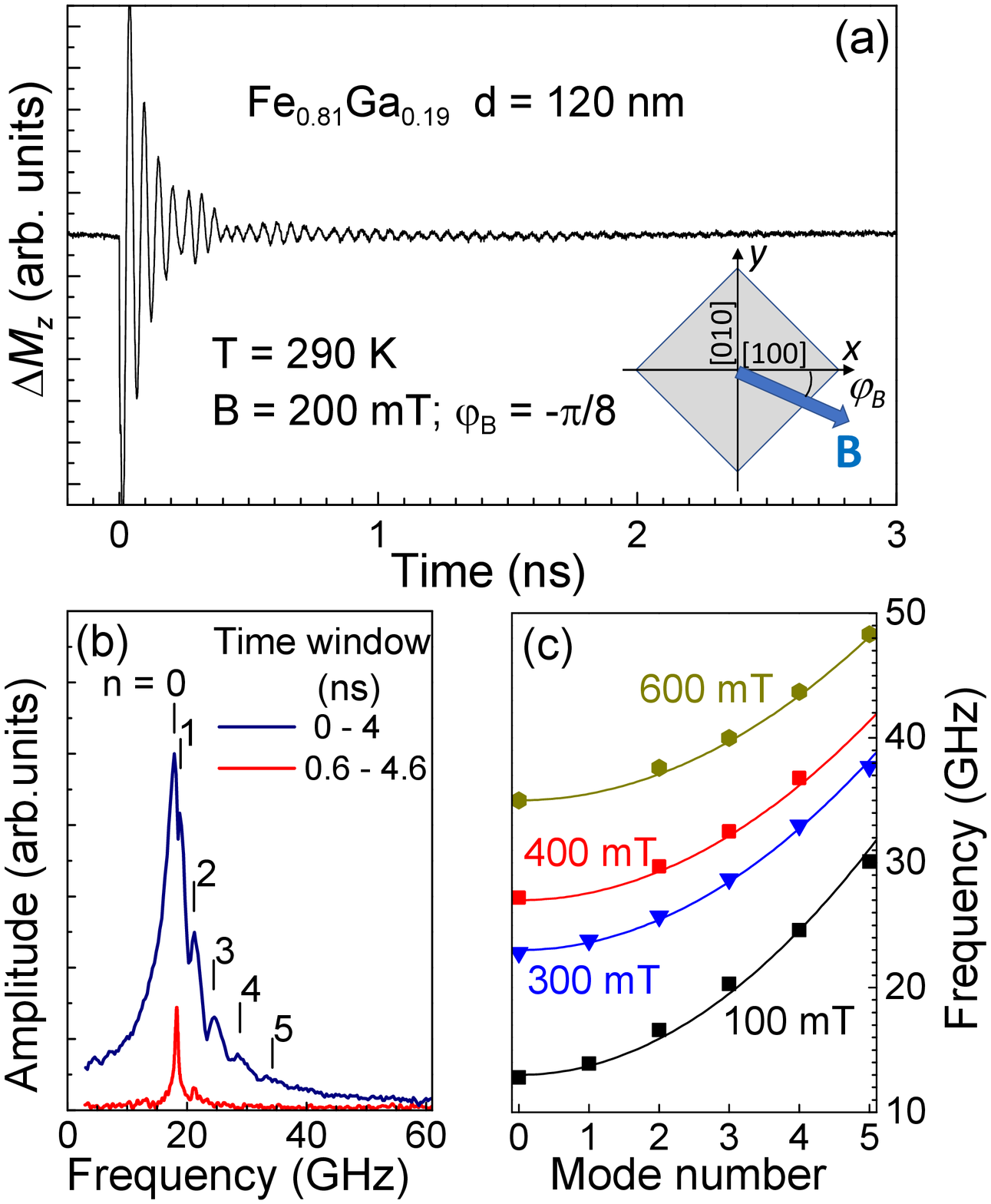}
 \caption{
Multimode magnon excitation; $T=290$~K. (a) Temporal evolution of the magnetization precession in a 120-nm thick Fe$_{0.81}$Ga$_{0.19}$ layer. (b) Fast Fourier transform of the signal shown in (a) performed in a time window of 4 ns, with the start point at $t=0$ (blue curve) and $t=0.6$ ns (red curve); vertical bars point at the frequency of resonance modes with $n=0,1,2,3\dots$(c) Measured (symbols) and calculated by Eq.(1) (lines) dependences of resonance frequency on the mode number for several in-plane magnetic fields. The inset in (a) shows the in-plane magnetic field configuration.
 }
\end{figure}

Figure 1 shows the experimental results for the thickest $d=120$~nm Fe$_{0.81}$Ga$_{0.19}$ layer obtained at $\varphi_B=-\pi/8$, when the precession amplitude is maximal. The magnetization precession shown in Fig. 1(a) decays in a time much less than 1 ns, which is consistent with the result for (Fe,Ga) films reported earlier \cite{Kats,Jasmin}. However, in contrast with the previous experiments, temporal beatings with a long-living tail are clearly observed. The fast Fourier transform (FFT) of the measured signal obtained in a time window of 4 ns is shown in Fig. 1(b). The blue line possesses a band spectrum where overlapping peaks are marked by integer numbers.  Six spectral bands with frequencies $f_n(n=0\dots 5)$ are recognized in the spectrum. We attribute these bands to standing spin wave (magnon) modes. This conclusion is based on a comparison of the experimental dependence of \textit{f${}_{n}$} on \textit{n,} shown in Fig. 1(c) by symbols, with the well-known dispersion relation for magnon modes:
\begin{equation}
f_{n} =f_{0} + \frac{1}{2\pi }\gamma_0\beta D q_n^2,
\end{equation}
where $q_n$ is the wavevector of the mode $n=0,1,2,3\dots$, $D$ is the exchange spin stiffness, $\gamma_0$ is the gyromagnetic ratio and $\beta$ is a field dependent coefficient determined by the anisotropy parameters of the ferromagnet \cite {Modes1}. With the assumption of free boundary conditions, $q_n=\pi n/d$, we get an excellent agreement of the measured magnon frequencies with the curves calculated using Eq.(1) for $D=2.6\times10^{-17}$~Tm$^2$, shown in Fig. 1(c) by lines \cite{Modes2}. This allows us to attribute unambiguously the bands in the measured spectra in this (Fe,Ga) film to magnon modes \cite{FeGa-magnons}.

 It is interesting that the FFT obtained in a temporal window which starts 600 ps after the pump pulse [red line in Fig. 1(b)] shows only two spectral lines with frequencies corresponding to $n=0$ and 2. We may conclude that different magnon modes have different decay times and that modes with uneven $n$ decay more quickly than modes with even $n$. The explanation of such behavior is related to the magnon decay mechanisms which are widely discussed in the literature \cite{DynamicsBook} but still not fully understood. Two-magnon scattering \cite{TwoMagnon} and the related selection rules could be the explanation, but this requires a comprehensive theoretical study which is beyond the scope of the present work.

\begin{figure}
 \includegraphics[scale=0.4]{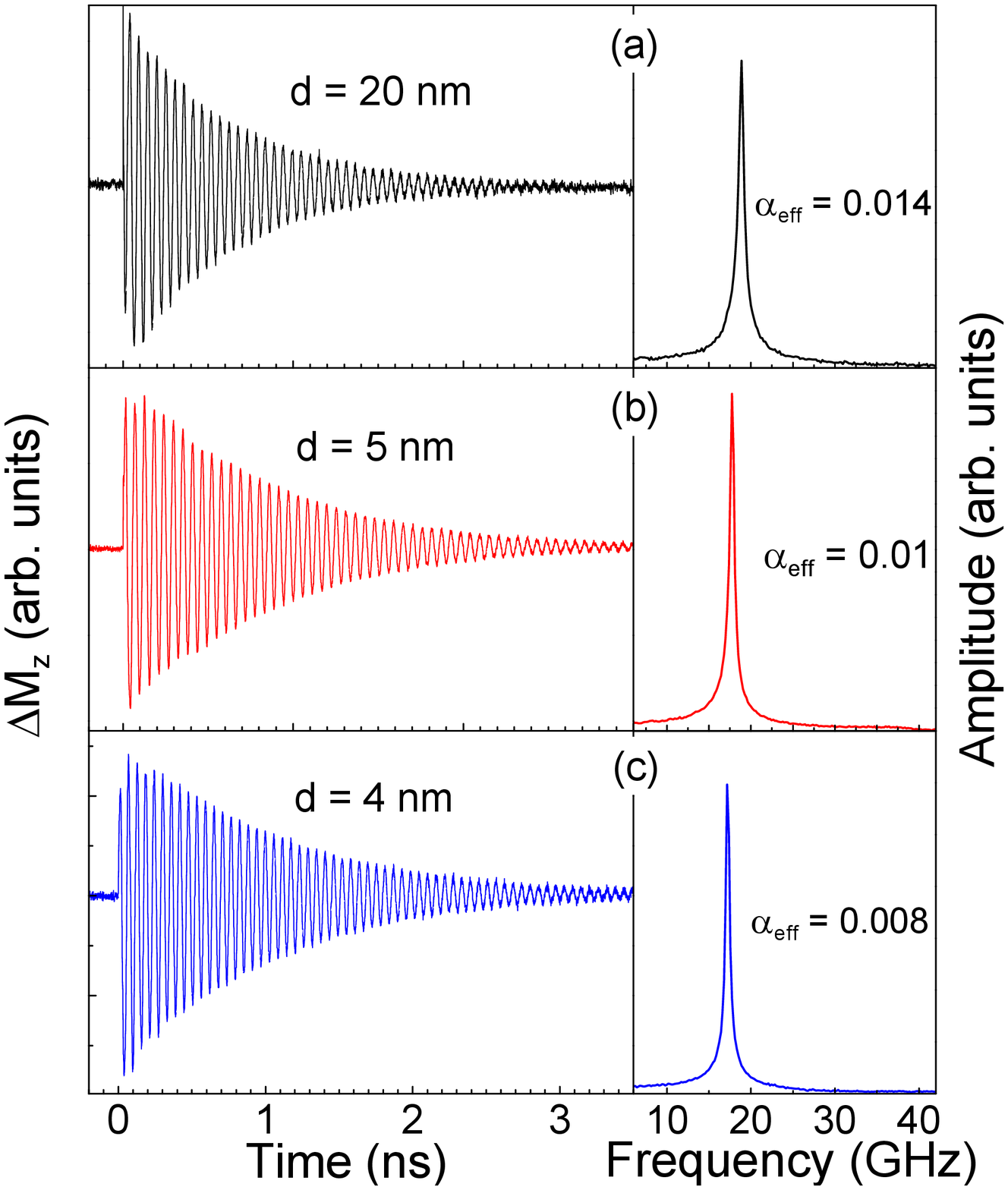}
 \caption{
Single-mode magnon excitation; $T = 290$~K. Temporal evolutions (left panels) and corresponding spectra (right panels) of the magnetization precession for Fe$_{0.81}$Ga$_{0.19}$ layers with different thickness measured at $B = 200$~mT and $\varphi_B=-\pi/8$.
 }
\end{figure}

The precession kinetics change drastically in thin nanolayers with $d=4$, 5 and 20 nm. Figure 2 shows the temporal evolutions (left panels) and their FFTs (right panels) of magnetization precession measured for $B=200$~mT applied at $\varphi_B=-\pi/8$. Only one spectral line is observed in the magnon spectrum, which corresponds to the fundamental mode with $n=0$.  The precession damping is well described with a single exponential decay with constants $\tau=1.05$, 0.85, and 0.6~ns, which correspond to $\alpha_{eff}=0.008$, 0.01 and 0.014 for the 4, 5, and 20-nm layers respectively.

Figure 3(a) shows the temporal evolution measured in the thinnest 4-nm nanolayer for $B=3$ T. The precession frequency is $f=108$ GHz, which corresponds to the maximum precession frequency in the present work. The FFT spectrum shown in the inset of Fig. 3(a) consists also of a Brillouin line at 44 GHz due to dynamical interference of the probe pulse on the strain pulse injected into the GaAs substrate \cite{Maris}, which is not related to the magnetic properties of the (Fe,Ga) layer. Single mode excitation is observed for the filtered signal (high-pass filter with 50 GHz cutoff frequency) shown in Fig. 3(b). The decay time of the magnetization precession in the 4 nm nanolayer at $f=108$ GHz is $\tau=0.29$~ns, which corresponds to an effective damping parameter $\alpha_{eff}=0.005$.

\begin{figure}
 \includegraphics[scale=0.48]{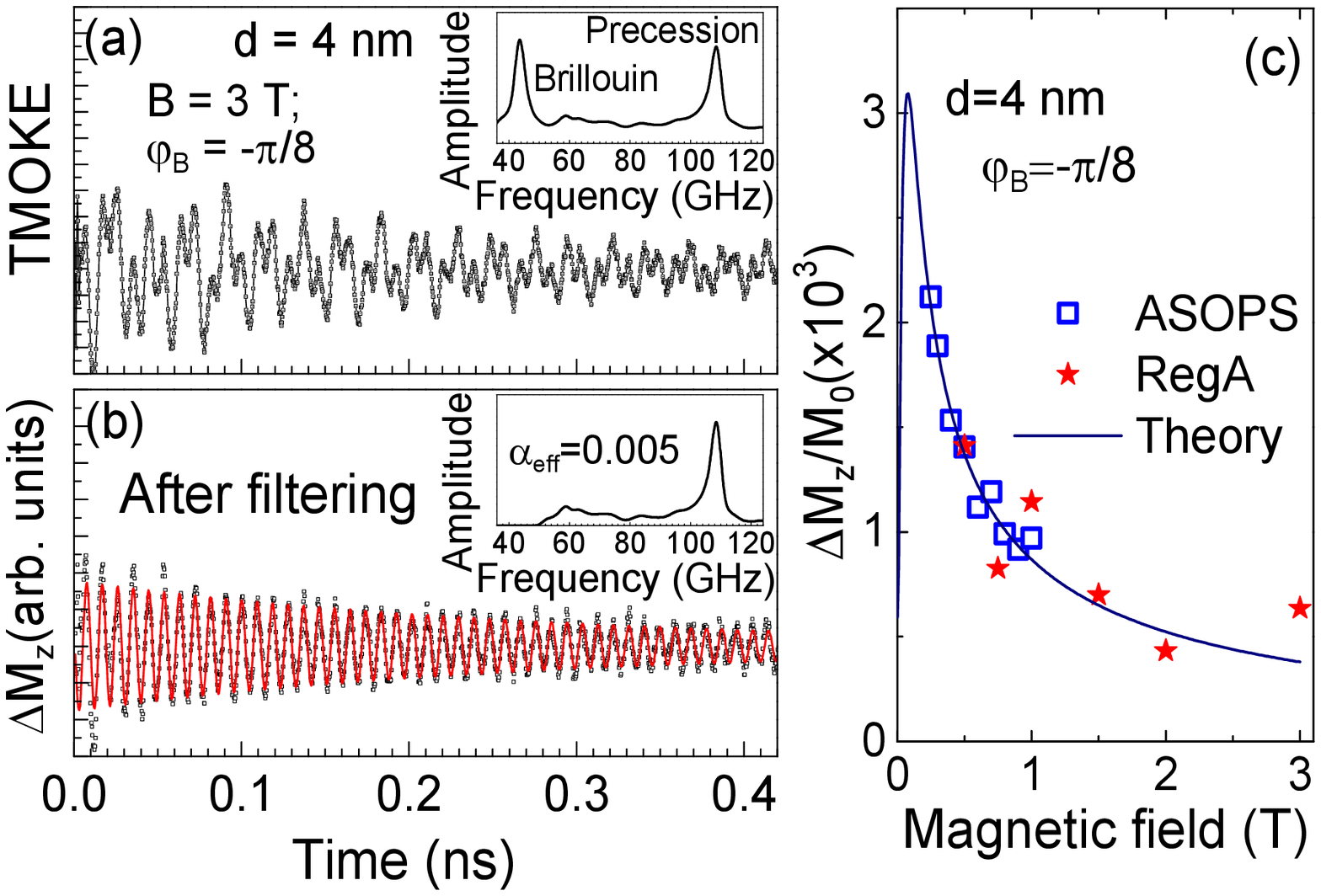}
 \caption{
(a) TMOKE signal and its FFT spectrum (inset) measured in 4-nm thick Fe$_{0.81}$Ga$_{0.19}$ nanolayer at $B=3$~T and $T=150$ K; (b) Temporal evolution of the magnetization precession obtained by high-pass filtering of the signal in (a); dots - experimental data; red line - fit with a single-frequency decaying sine function; inset is the corresponding FFT spectrum; (c) Normalized experimental (symbols) and calculated (solid line) dependences of the precession amplitude on external magnetic field; squares and stars correspond to the data measured by the ASOPS system and by scanning delay line with excitation by the RegA, respectively.
 }
\end{figure}

The line in Fig. 3(b) is a fit to the experimental data by an exponentially decaying sine function:
\begin{equation}
\Delta M_z=A\exp{(-t/\tau)}\sin{(\omega t+\psi)},
\end{equation}
where $\omega=2 \pi f$ ($f$ is obtained from the FFT spectrum). The fitting parameters $A$, $\tau$, and $\psi$ are the amplitude, decay time and the initial precession phase, respectively. The dependence of the amplitude, $A$ on $B$ for the thinnest (Fe,Ga) nanolayer is shown by symbols in Fig. 3(c). It is seen that $A$ decreases with increasing $B$, but in our experiment it is still possible to detect the precession with frequency higher than 100 GHz at $B=3$~T.

The main experimental results of the present work are the demonstration of excitation of a multimode quantized precession spectrum in a thick, 120-nm, (Fe,Ga) layer, and a long-living single mode magnetization precession with a frequency $>100$~GHz in a thin, 4-nm, (Fe,Ga) nanolayer. Our qualitative explanation for these experimental facts is based on a comparison of the optical penetration depth in (Fe,Ga) with the layer thickness, $d$. The penetration depth for the pump light is $\eta \approx 20$~nm, which is larger than the thickness of the films where only one magnon mode is excited. In this case, the optical excitation, which kicks the magnetization precession, is almost homogeneous along the thickness of the nanolayer. Assuming free boundary conditions at the nanolayer interfaces, only the excitation of the ground uniform mode is efficient, while the higher order magnon modes are not excited due to their sign-changing spatial profile \cite{byLaser}. In contrast, in thick films $\eta<d$, and the excitation is inhomogeneous, being stronger near the surface, resulting in the efficient excitation of high-energy magnon modes. The efficiency of such excitation should decrease with the increase of $n$, which is clearly observed in Fig. 1(b): the spectral amplitude of the magnon spectral line decreases by more than one order of magnitude with $n$ increasing from 0 to 5. It is important to note that due to the shallow penetration depth of the probe pulse, both even and odd magnon modes contribute to the TMOKE signal and we observe monotonic decrease of the magnon mode amplitude with increase of its number.

 We now consider the observation of precession with frequency $\approx100$ GHz.  Fitting the measured temporal signal shown in Fig. 3(b) with a single harmonic function gives a decay $\tau$=0.29 ns and a respective value for $\alpha_{eff}=0.005$. This value is close to the smallest damping parameters measured in pure Fe on semiconductor substrates by the FMR technique \cite{FMR1,FMR2,FMR3,FMR4}, but has not been reported in experiments using ultrafast optical excitation of the magnetization precession in metallic ferromagnetic materials so far.

We have performed a theoretical analysis of the precessional response of the magnetization and its dependence on magnetic field strength and direction using the approach presented in earlier work \cite{Kats}, which considers launching of the magnetization precession by ultrafast modification of the magnetic anisotropy. The comprehensive study of the angular dependences $f(\varphi_B)$ and $A(\varphi_B,B)$, which can be found in the Supplemental Material \cite{Supplement}, allows us to obtain the main film parameters: saturation magnetization $\mu_0M_0=1.72$ T, cubic anisotropy coefficient $K_1=15$ mT and uniaxial anisotropy coeficient $K_u=5$ mT. We also confirmed experimentally that for the used pump excitation density, the demagnetization is negligible \cite{Supplement}. The optically-induced changes of the anisotropy coefficients were estimated by using the data from Ref. \cite{Kats}: $\Delta K_1=-3.7$ mT and $\Delta K_u=-0.8$ mT. The respective dependence of the precession amplitude on magnetic field calculated at $\varphi_B=-\pi/8$ is shown by the solid line in Fig. 3(c). A good agreement between the experimental dependence, which is normalized accordingly, and the theoretical curve is clearly observed. Moreover, for relatively small changes of the anisotropy coefficients, and neglecting demagnetization, we can simplify the dependence $A(\varphi_B,B)$ to:

\begin{equation}
A\approx \frac{(\Delta K_1/2)\sin{4\varphi_B}-\Delta K_u\cos{2\varphi_B}}{\sqrt{B(B+\mu_0 M_0)}}.
\end{equation}
This expression is valid with high accuracy at $B\geq200$~mT. As one can see from Eq.(3), the precession amplitude is maximal at $\varphi_B=-\pi/8$ ($-3\pi/8$, $5\pi/8$, and $7\pi/8$), and remains nonzero with increase of magnetic field due to the field-independent $\Delta K_1$ and $\Delta K_u$. At $B=9$~T, when the precession frequency approaches the terahertz range ($f=300$~GHz), the estimated precession amplitude $\Delta M_z/M_0=3\times10^{-4}$ is expected to be easily detectable.

It is worth noting that in the 4-nm layer $\alpha_{eff}$ demonstrates a pronounced anisotropy and is 1.5 times smaller at $\varphi_B=\pi/4$ than at $\varphi_B=-\pi/8$ \cite{Supplement}. Unfortunately, the small precession amplitude at $\textbf{B}\parallel[110]$ does not allow us to detect the magnetization precession at high magnetic fields applied along this direction. Anisotropic damping has been previously observed in Fe nanolayers and is actively studied nowadays \cite{FMR2,FMR3,FMR4}.

In conclusion, we have demonstrated multimode excitation of magnetization precession in Fe$_{0.81}$Ga$_{0.19}$ layers with a thickness of 120 nm and single-mode precession in thin Fe$_{0.81}$Ga$_{0.19}$ nanolayers. We show that the parameters of (Fe,Ga) provide the possibility to detect magnetization precession with frequency higher than 100 GHz, and small effective damping parameter $\alpha_{eff}\approx0.005$. These are record values for experiments using optical excitation of magnetization precession in metallic ferromagnets. Due to the large saturation magnetization, the precession amplitude of $10^{-3}M_0$ observed at high magnetic fields generates an ac-induction of ~1 mT, which may be exploited for nanoscale generators of microwave magnetic field \cite{Grating} and pure spin currents \cite{Pumping}. Our analysis shows that 100 GHz is not the limit for the detectable magnetization precession and the THz range can be achieved by applying an appropriate external magnetic field.

\section{Acknowledgements}
We are grateful to Serhii Kukhtaruk and Alexandra Kalashnikova for fruitful discussions. This work was supported by the Deutsche Forschungsgemeinschaft and the Russian Foundation for Basic Research in the frame of the International Collaborative Research Center TRR160 [project B6] and by the Bundesministerium f\"{u}r Bildung und Forschung through the project VIP+ "Nanomagnetron". The experimental studies in the Laboratory of Physics of Ferroics (Ioffe Institute) were performed under support of the Russian Science Foundation [grant no. 16-12-10485]. The Volkswagen Foundation supported the cooperation with the Lashkarev Institute [grant no. 90418].


\begin{thebibliography}{99}
\bibitem{Modes1} We use a simplified formula for the spin wave dispersion (the first two terms of the Taylor series of the dispersion relation, which includes the magneto-crystalline and shape anisotropies - see Ref. \cite{Gurevich} for details). If external magnetic field is significantly larger than the anisotropy field(s), $\beta=1$. The best fit of the experimental data gives:  $\beta=1.45$, 1.22, 1.15 and 1.05 for $B=100$, 300, 400 and 600 mT, respectively.
\end{thebibliography}
\end{document}